\documentclass[prb,aps,twocolumn,showpacs,amsmath,amssymb,floatfix,citeautoscript,10pt]{revtex4-1}
\usepackage{graphicx}
\usepackage{amsmath}
\usepackage{amssymb}
\usepackage{times}
\usepackage{epsfig,subfigure}
\usepackage{color}
\usepackage{float}
\usepackage{hyperref}

\newcommand {\vri} {{\bf R}} 
\newcommand {\vrri} {{\bf R}'} 
\newcommand {\sa} {\sigma} 
 
\newcommand {\vk} {{\bf k}}

\newcommand {\vq} {{\bf Q}}

\newcommand \vqaf {\vq_{\text{AF}}}
\newcommand \med[1] {\langle{#1}\rangle}

\hypersetup{
pdfauthor = {S\'eb,Chris and Cath},
pdftitle = {Spin liquid versus long range magnetic order in the frustrated body-centered tetragonal lattice. \today},
}

\begin{document}

\title{Spin liquid versus long range magnetic order in the frustrated body-centered tetragonal lattice}

\author{Carlene Farias}
\affiliation{Univ. Bordeaux, LOMA, UMR 5798, F-33400 Talence, France}
\affiliation{CNRS, LOMA, UMR 5798, F-33400 Talence, France}
\affiliation{International Institute of Physics, Universidade Federal do Rio Grande do Norte, 59078-400 Natal-RN, Brazil}
\affiliation{Departamento de F\'isica Te\'orica e Experimental, Universidade Federal do Rio Grande do Norte,  59072-970 Natal-RN,Brazil}
\author{Christopher Thomas}
\affiliation{Instituto de F\'isica, Universidade Federal do Rio Grande do Sul, 91501-970 Porto Alegre-RS, Brazil}
\author{Catherine P\'epin}
\affiliation{Institut de Physique Th\'eorique, CEA-Saclay, 91191 Gif-sur-Yvette, France}
\author{Alvaro Ferraz}
\affiliation{International Institute of Physics, Universidade Federal do Rio Grande do Norte, 59078-400 Natal-RN, Brazil}
\affiliation{Departamento de F\'isica Te\'orica e Experimental, Universidade Federal do Rio Grande do Norte,  59072-970 Natal-RN,Brazil}
\author{Claudine Lacroix}
\affiliation{Institut N\'eel, Universit\'e Grenoble-Alpes, F-38042 Grenoble, France}
\affiliation{Institut N\'eel, CNRS, F-38042 Grenoble, France}
\author{S\'{e}bastien Burdin}
\affiliation{Univ. Bordeaux, LOMA, UMR 5798, F-33400 Talence, France}
\affiliation{CNRS, LOMA, UMR 5798, F-33400 Talence, France}

\begin{abstract}
An $SU({\rm n})$-symmetric generalization of the Heisenberg model for quantum spin $S$ operators is used to investigate 
the geometrically frustrated body-centered tetragonal (BCT) lattice with antiferromagnetic interlayer coupling $J_1$ and intralayer first and second neighbor coupling $J_2$ and $J_3$. Using complementary representations of the spin operators, we study the phase diagram characterizing the ground state of the system. For small n, we find that the most stable solutions correspond to four different families of possible long range magnetic orders that are governed by $J_1$, $J_2$, and $J_3$. 
First, some possible instabilities of these phases are identified for $n=2$ in large $S$ expansions up to the linear spin-wave corrections. Then, using a fermionic representation of the $SU({\rm n})$ spin operators for $S=1/2$, we find that purely 
magnetic orders occur for ${\rm n}\le 3$ while spin-liquid (SL) solutions are stabilized for ${\rm n}\ge 10$. 
The SL solution governed by $J_1$ breaks the lattice translation symmetry. This Modulated SL is associated to a commensurate ordering wave vector $(1,1,1)$. For $4\le {\rm n}\le 9$, we show how competition between $J_1$, $J_2$, and $J_3$ can tune the ground state from beeing magnetically ordered to a SL state. We discuss the relevance of this scenario for correlated systems with BCT crystal structure. 
\end{abstract}

\date{\today}

\maketitle
{\it Introduction.---}The body-centered tetragonal (BCT) lattice is one of the 14 three-dimensional (3D) lattice types~\cite{Kittelbook}. 
This standard crystalline structure is realized in several strongly correlated electron materials with unusual magnetic 
and transport properties. Among the heavy fermion systems~\cite{Stewart1984, Fulde2006}, 
different examples of materials with rare earth atoms on a BCT lattice have been intensively studied for the last decades: 
in URu$_2$Si$_2$, a still mysterious Hidden order (HO) phase appears below the critical temperature $T_{HO}\approx 17~{\rm K}$ close to a 
pressure-induced antiferromagnetic 
(AF) transition~\cite{Palstra1985, Mydosh2011}; in YbRh$_2$Si$_2$ and CeRu$_2$Si$_2$, non-Fermi liquid properties  are observed in the vicinity of AF quantum phase transitions, that are still poorly 
understood~\cite{Steglich2003,Steglich2009, Flouquet1988, Knafo2009}; 
CeCu$_2$Si$_2$ was the first heavy fermion material where unconventional superconductivity was discovered close to an AF transition~\cite{Steglich1979}; CePd$_2$Si$_2$ also exhibits unconventional superconductivity related to an AF transition~\cite{Lonzarich1998, Demuer2001}; multi-{\bf Q} AF order has been observed in CeRh$_2$Si$_2$~\cite{Kawarasaki2000}. 
Today, each one of those "122" compounds can yet be considered as one entire field of research. 
It is noticeable that the link between AF ordering and unconventional superconductivity has also been suggested in other families of correlated materials with BCT symmetry. In particular, the cuprate superconductors~\cite{Bednorz1986} include among the AF insulating parent compounds La$_2$CuO$_4$ and 
Sr$_2$CuO$_2$Cl$_2$ in which the AF order originates from the Cu atoms that form a BCT crystal. 
However the relevant physics in their case is essentially two-dimensional (2D) with the BCT structure being involved only in the formation of 
the square-lattice layers of Cu atoms that order antiferromagnetically. 
Most of these materials are metals or superconductors, but their unconventional properties are strongly related to the interplay  between charge and magnetic degrees of freedom. In this letter, our approach to these systems with itinerant electrons is orthogonal to several standard approaches since we will start from a localized point of view. We suggest that the rich diversity of unusual physical properties observed in these materials with BCT-structure is associated with the diversity of the underlying magnetic phases in competition. 

Important theoretical developments were made in the past years on the unconventional magnetic properties of the BCT lattice using a classical Heisenberg model, following the pioneering study by 
Villain~\cite{Villain1959,Diep1989,Diep1989a,Rastelli1989,Rastelli1990,Quartu1998,Loison2000,Sorokin2011}. 

\begin{figure}[H]
\centering
\includegraphics[width=0.9\columnwidth]{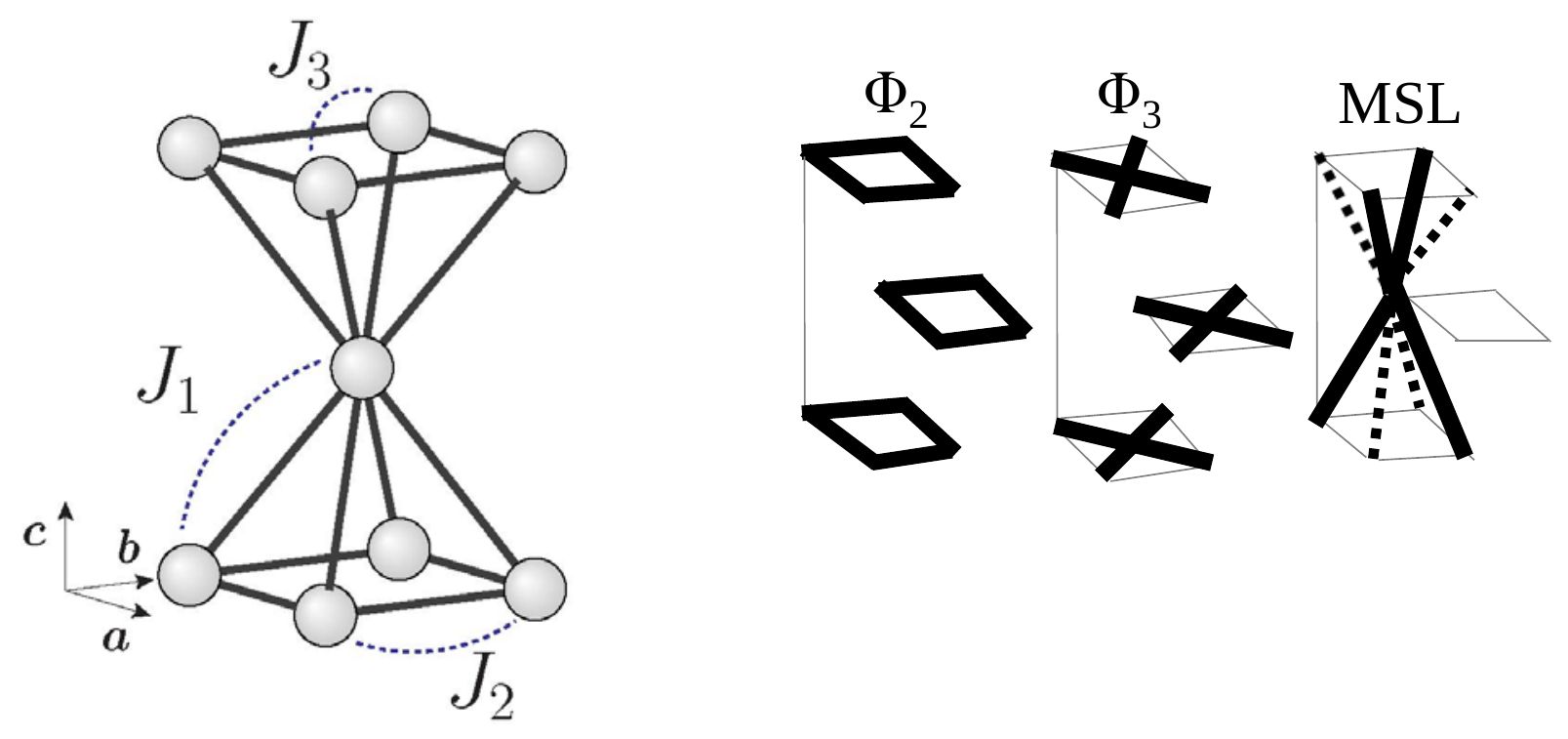}
\caption{\label{fig:bct} Left: BCT lattice and the $J_1$, $J_2$, and $J_3$ interactions. 
In this letter, the tetragonal lattice constants are set to $a=b=c=1$. Right: bold lines represent the three kinds of intersite SL correlations on the BCT structure. }
\end{figure}
{\it Model.---}In this letter, we analyze the ground states of an $SU({\rm n})$ generalization of the
$J_1$-$J_2$-$J_3$ quantum Heisenberg Hamiltonian introduced here first for ${\rm n}=2$: 
\begin{eqnarray}
H_{{\rm n}=2}^{S}=\sum_{\langle {\bf R,R'}\rangle} J_{\bf RR'}\vec{S}_{\bf R}\cdot\vec{S}_{\bf R'}~, 
\label{HamiltonianSU2}
\end{eqnarray}
where $\vec{S}_{\bf R}\equiv (S_{\bf R}^{x}, S_{\bf R}^{y}, S_{\bf R}^{z})$ denotes quantum spin $S$ operators acting on site ${\bf R}$ of a BCT-lattice. The antiferromagnetic interaction $J_{{\bf RR}'}$ connects sites 
${\bf R}$ and ${\bf R}'$, and can take three possible values $J_1,~J_2,~J_3>0$, as indicated in 
figure~\ref{fig:bct}. 
\begin{figure}[H]
\centering
\includegraphics[width=\columnwidth]{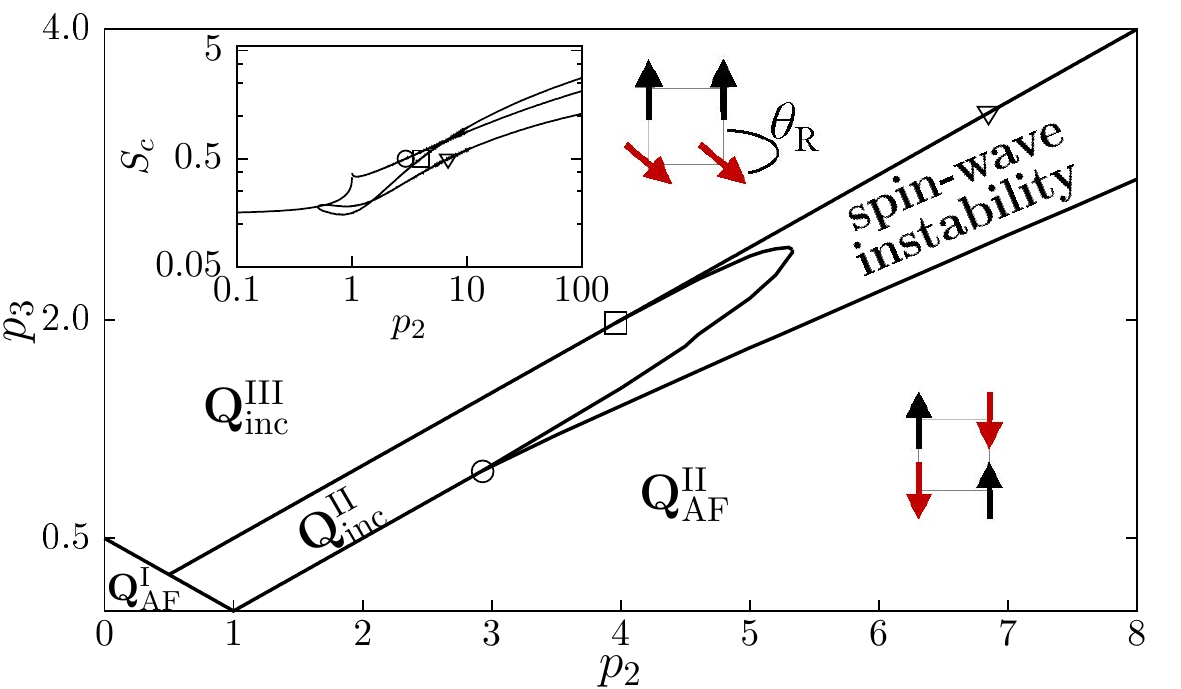}
\caption{(Color online) Classical ground state phase diagram of the $J_1$-$J_2$-$J_3$ model in coordinates 
$(p_2, p_3)$. The solid lines indicate transitions between the stable magnetic ordered states with modulating vectors 
${\bf Q}_{\rm AF}^{\rm I}$, ${\bf Q}_{\rm AF}^{\rm II}$, ${\bf Q}_{\rm inc}^{\rm II}$, 
and ${\bf Q}_{\rm inc}^{\rm III}$. A schematic representation is provided for ${\bf Q}_{\rm AF}^{\rm II}$ and 
${\bf Q}_{\rm inc}^{\rm III}$. 
The linear SW instability region is depicted here for $S=1/2$. 
Inset: critical value $S_c$ as a function of $p_2$ when $p_3$ is bound to the critical lines. 
The three symbols indicate the corresponding location of critical value $S_c=1/2$ in each case. 
\label{fig:classical} }
\end{figure}
{\it Classical ground state.---}Following the standard spin-wave (SW) approach, we start with the $S=+\infty$ generalization of the Hamiltonian~(\ref{HamiltonianSU2}), which corresponds to the limit of classical spins. For the sake of simplification, we 
consider only magnetic orderings that are characterized by a single wavevector 
${\bf Q}=2\pi(\lambda, \mu, \nu)$ identified as $(\lambda,\mu, \nu)$ in reduced notations. 
Invoking Fourier transforms, this wavevector is used to minimize the classical dispersion
$J({\bf q})\equiv 8 J_1\gamma_{1}^{\bf q}+2J_2\gamma_{2}^{\bf q}
+4J_3\gamma_{3}^{\bf q}$, with
\begin{eqnarray}
\gamma_{1}^{\bf q}&\equiv&\cos{(q_x/2)}\cos{(q_y/2)}\cos{(q_z/2)}\,,\label{defgamma1k}\\
\gamma_{2}^{\bf q}&\equiv&\cos{(q_x)}+\cos{(q_y)}\,,\label{defgamma2k}\\
\gamma_{3}^{\bf q}&\equiv&\cos{(q_x)}\cos{(q_y)}\,.\label{defgamma3k}
\end{eqnarray}
Tuning the dimensionless parameters $p_2\equiv J_2/J_1$ and $p_3\equiv J_3/J_1$, we find that the ground state can be characterized by four kinds of possible wavevectors, as depicted in figure~\ref{fig:classical}: 
${\bf Q}_{\rm AF}^{\rm I}\equiv (1,1,1)$ and 
${\bf Q}_{\rm AF}^{\rm II}\equiv (1/2, 1/2, \nu)$ correspond to the regimes where the Weiss field is dominated by 
$J_1$ and $J_2$ respectively. The $\nu$-degeneracy in the latter case indicates the underlying bidimentionality. 
The other possible ordering wavevectors are incommensurate and characterize two kinds of helical orders: 
${\bf Q}_{\rm inc}^{\rm III}\equiv (0,{\Upsilon_3},0)$ degenerate with $({\Upsilon_3},0,0)$, $(1,{\Upsilon_3},1)$, 
and $({\Upsilon_3},1,1)$ where ${\Upsilon_3}=\frac{1}{\pi}\arccos{\frac{-1}{p_2 + 2p_3}}$; and 
${\bf Q}_{\rm inc}^{\rm II}\equiv (\Upsilon_2, \pm\Upsilon_2, 1)$ degenerate with 
$(\Upsilon_2, 1\pm\Upsilon_2, 0)$, where ${\Upsilon_2}=\frac{1}{2\pi}\arccos{\frac{1-p_2}{2p_3}}$. 
A different wave-vector, ${\bf Q}_{\rm AF}^{\rm III}\equiv (0, 1/2, \nu)$ had been proposed~\cite{SUGIYAMA1990} in a $J_3$-dominated phase, which corresponds to the commensurate order characterizing a purely bidimensional square lattice. 
We find that $J_1\neq 0$ corrections are relevant and ${\bf Q}_{\rm inc}^{\rm III}$ is energetically more stable than 
${\bf Q}_{\rm AF}^{\rm III}$. Not surprisingly these two vectors are asymptotically identical at large $p_3$. 
Similarly, ${\bf Q}_{\rm inc}^{\rm II}\mapsto{\bf Q}_{\rm AF}^{\rm II}$ at large $p_2$. 
The transition lines separating these four classical phases are given by linear relations between $p_2$ and $p_3$. 
The ${\bf Q}_{\rm inc}^{\rm II}-{\bf Q}_{\rm inc}^{\rm III}$ transition is dicontinuous, the other transitions are continuous. 

{\it Order by quantum disorder.---}Then, assuming a given classical ground state ${\bf Q}$, we study the large-$S$ corrections. 
This expansion invokes a helical generalization of the Holstein-Primakov representation~\cite{Chubukov1984,Diep1989, Rastelli1990}: introducing boson anihilation (creation) $a_{\bf R}^{(\dagger)}$, spin operators are approximated as 
$S_{\bf R}^{x}\approx(S/2)^{\frac{1}{2}}(a_{\bf R}+a_{\bf R}^{\dagger})$, and 
$\left(\begin{array}{c}S_{\bf R}^{z} \\ S_{\bf R}^{y}\end{array}\right)\approx \left[
\begin{array}{rl}\cos\theta_{\bf R}&\sin\theta_{\bf R}\\-\sin\theta_{\bf R}&\cos\theta_{\bf R}\end{array}\right]
\left(\begin{array}{l}S-a_{\bf R}^{\dagger}a_{\bf R} \\ -i(S/2)^{\frac{1}{2}}(a_{\bf R}-a_{\bf R}^{\dagger})
\end{array}\right)$, with $\theta_{\bf R}\equiv{\bf Q}\cdot({\bf R}-{\bf R}_0)$, where 
$z$ is the easy axis characterizing a site ${\bf R}_0$ choosen arbitrarily. 
The dispersions obtained for the Bogoliubov quasiparticles are 
$\Omega_{\bf k}^{+}\equiv J({\bf k})-J({\bf Q})$ and 
$\Omega_{\bf k}^{-}\equiv \frac{J({\bf k}+{\bf Q})+
J({\bf k}-{\bf Q})}{2}-J({\bf Q})$. 

Whilst the ground state energy is proportional to $S^2J({\bf Q})$ at the highest order, the first correction is proportional to 
$S\int_{\rm BCT}d^3{\bf k}\sqrt{\Omega_{\bf k}^{+}\Omega_{\bf k}^{-}}$ where the ${\bf k}$-integral runs over the first Brillouin zone of the BCT-lattice. 
Analyzing this correction for the classicaly degenerate vectors ${\bf Q}_{\rm AF}^{\rm II}$, we find that 
the resulting order is stabilized by quantum disorder: the continuous degeneracy is lift in favor of $\nu=0$, which is equivalent to $\nu=1$. 

{\it Fluctuation corrections to magnetization.---}We also studied the effects of fluctuations emerging from the linear 
SW corrections. 
Generalizing to the BCT-structure the approach introduced in~\cite{Chandra1988} for the square lattice model,  
the staggered magnetization is expanded around its classical value $\langle S_{{\bf R}_0}^{z}\rangle\approx S-\Delta m(p_2, p_3)$. We find: 
\begin{eqnarray}
\Delta m(p_2, p_3)= \langle a_{\bf R_0}^{\dagger}a_{\bf R_0}\rangle
=-\frac{1}{2}+\int_{\rm BCT}\frac{d^3{\bf k}}{64\pi^{3}}
\frac{\Omega_{\bf k}^{+}+\Omega_{\bf k}^{-}}{\sqrt{\Omega_{\bf k}^{+}\Omega_{\bf k}^{-}}}~, \nonumber\\
~
\end{eqnarray}

Unlike the 2D case~\cite{Chandra1988}, fluctuation corrections here do not diverge, which is not surprising with a 
3D model. Frustration can relatively increase the critical value of $S$ below which the linear spin-wave correction 
cancels the staggered magnetization, $S_c\equiv\Delta m(p_2, p_3)$. 
Indeed, for fixed $p_2$, we find that $S_c$ increases when $p_3$ approaches its critical value associated with the 
classical phase boundary. This maximal value is plotted in the inset of figure~\ref{fig:classical} as a function of $p_2$ 
for the continuous ${\bf Q}_{\rm AF}^{\rm II}/{\bf Q}_{\rm inc}^{\rm II}$ transition and on each side of the 
discontinous ${\bf Q}_{\rm inc}^{\rm II}/{\bf Q}_{\rm inc}^{\rm III}$ transition. 
On each of these critical lines, we find $S_c\sim\sqrt{p_2}$ at large $p_2$. 
Furthermore, a logarithmic $S_c\sim \ln(p_2/2 -p_3)$ and a power law $S_c\sim 1/\sqrt{p_3-p_2/2}$ are respectively 
obtained at large $p_2$ in the vicinity of the ${\bf Q}_{\rm inc}^{\rm II}\to{\bf Q}_{\rm inc}^{\rm III}$ 
and ${\bf Q}_{\rm inc}^{\rm III}\to{\bf Q}_{\rm inc}^{\rm II}$ transitions. This result is consistent with the square lattice spin wave analysis~\cite{Chandra1988}. 

{\it Generalized $SU({\rm n})$ symmetric model.---}The SW approach thus reveals some weaknesses of the classical magnetic orders, but the three-dimensionality 
protects these states against small fluctuations at the lowest, linear, order. 
One may go further and study possible instabilities emerging from next orders, taking into account 
interactions between the spin-wave bosonic excitations. 
Hereafter, we follow an alternative approach: we analyze the possibility that the system forms a 
Resonant Valence Bond state with fermionic excitations and SL correlations~\cite{Aharony1976, Fazekas1974}. 
One of our physical motivations is driven by the physics of unconventional metallic systems with BCT 
structure: in several of these correlated systems, magnetic degrees of freedom seem to be "deconfined" into fermionic ones that may contribute to the formation of a Fermi-surface, unlike weakly coupled bosons. Such a scenario inspired by the 
physics of cuprate superconductors~\cite{Anderson1987, Baskaran1987, Rice1993, Wen1996} could be easily 
strengthened by a coupling of the Heisenberg spins of the $J_1$-$J_2$-$J_3$ model to extra charge degrees of freedom. 
In the following we will study this possible "fermionic deconfinement" of spin operators as an intrinsic property of the 
Heisenberg model. To this goal, considering $S=1/2$, the Hamiltonian~(\ref{HamiltonianSU2}) is generalized to 
$SU({\rm n})$-symmetry: 
\begin{eqnarray}
H^{S=1/2}_{\rm n}=\sum_{\langle {\bf R},{\bf R'}\rangle}\frac{J_{{\bf RR}'}}{\rm n}
\sum_{\sigma\sigma'}\chi_{{\bf R}\sigma}^{\dagger}\chi_{{\bf R}\sigma'}\chi_{{\bf R}'\sigma'}^{\dagger}\chi_{{\bf R}'\sigma}~, 
\label{eq:hheis}
\end{eqnarray}
where $\chi_{{\bf R}\sigma}^{(\dagger )}$ ($\chi_{{\bf R}\sigma}$) are annihilation (creation) fermionic operators 
with orbital degeneracy $\sigma=1,\cdots, {\rm n}$, and satisfying the local constraints 
$\sum_{\sigma}\chi_{{\bf R}\sigma}^{\dagger}\chi_{{\bf R}\sigma}={\rm n}/2$. This is a standard $SU({\rm n})$ 
generalization~\cite{Affleck1988} of the fermionic representation developped by Abrikosov for ${\rm n}=2$.  The scaling factor $1/{\rm n}$ ensures that the energy 
remains extensive, i.e., proportional to ${\rm n}$, in the large-${\rm n}$ limit. 

{\it Spin-liquid correlations.---}Using the Hubbard-Stratonovitch decoupling as described in~\cite{Pepin2011,Thomas2013}, 
the low temperature phases of Hamiltonian~\eqref{eq:hheis} can be characterized by 
two kinds of order parameters: the local magnetization field 
$m_{\bf R}^{\sigma}=\langle \chi_{\vri\sa}^{\dag}\chi_{\vri\sa}\rangle-\frac{1}{2}$, and the 
intersite spin-liquid fields 
$\varphi_{\vri\vrri}=-\frac{1}{\rm n}\sum_{\sigma}\med{\chi_{\vri\sa}^{\dag}\chi_{\vrri\sa}}$. 
The purely magnetic classical mean-field theory characterized here by a staggered magnetization 
$m_{{\bf R}}^{\sigma}=\pm S_{\bf Q}$ is equivalent to the one we analyzed for the $S=+\infty$ limit. The corresponding ground state phase diagram is thus given by figure~\ref{fig:classical}. Hereafter, the stability of these classical 
magnetic orders is analyzed by testing various SL Ans\"atze as alternative possible ground states. Generalizing the Modulated SL (MSL) order introduced in~\cite{Thomas2013}, we consider the nearest neighbor intersite correlations $\varphi_{\vri\vrri}^{1}=\frac{1}{2}\Big[\Phi_1+ie^{i{\bf Q}_{\rm AF}^{\rm I}
\cdot\big(\frac{\vri+\vrri}{2}\big)}\Phi_{\rm M}\Big]$, $\varphi_{\vri\vrri}^{2}=\Phi_2$, and $\varphi_{\vri\vrri}^{3}=\Phi_3$ with a bond index definition similar to the one of figure~\ref{fig:bct}. 
The free energy per spin component and per lattice site is expressed for each AF or SL state as: 
\begin{eqnarray}
F=F_0-\frac{k_B T}{32\pi^3}\int_{\rm BCT} d^3{\bf k}\sum_{s=\pm}\ln{
\left(1+e^{-\frac{E_{\vk}^{s}}{k_B T}}\right)}-\frac{\lambda_{0}}{2}~, 
\end{eqnarray}
where $\lambda_0$ denotes a Lagrange multiplier 
that minimizes $F$ in order to satisfy the constraint for the fermionic occupation. 
For the AF orderings, the zero-point energy is $F_0=-\frac{2J({\vq})}{n}\vert S_{\vq}\vert^{2}$ 
and the dispersion $E_{\vk}^{\pm}=\lambda_0\pm \frac{J({\vq})}{n}S_{\vq}$. 
For the SL states we find 
$F_0/J_{1}=\vert\Phi_{1}\vert^{2}+\vert\Phi_{\rm M}\vert^{2}
+2p_{2}\vert\Phi_{2}\vert^{2}+2p_{3}\vert\Phi_{3}\vert^{2}$ and 
$(E_{\vk}^{\pm}-\lambda_0)/J_1=2p_2\gamma_{2}^{\vk}\Phi_2
+4p_3\gamma_{3}^{\vk}\Phi_3
\pm 4 
\sqrt{
(\gamma_{1}^{\vk}\Phi_{1})^{2}+(\gamma_{\rm M}^{\vk}\Phi_{\rm M})^{2}}$. 
Here, the non-BCT-periodic real term $\Phi_{\rm M}$ with 
$\gamma_{\rm M}^{\vk}\equiv 
\sin{(q_x/2)}\sin{(q_y/2)}\sin{(q_z/2)}$ takes into account a possible spatial amplitude modulation of the SL field. 
Considering that this MSL is a BCT adaptation of the "kite" phase studied in~\cite{Affleck1988} for a square lattice, 
we also tested another non-homogeneous "flux" phase SL characterizing a chiral state with complex 
$\varphi_{\vri\vrri}^{1}=(\varphi_{\vrri\vri}^{1})^\star$ that could be described within a very close formalism by simply 
replacing $\gamma_{\rm M}^{\vk}\to\sin{(q_x/2)}\sin{(q_y/2)}\cos{(q_z/2)}$. This chiral SL 
was found to have a higher energy than the MSL.

\begin{figure}[H]
\centering
\includegraphics[width=0.9\columnwidth]{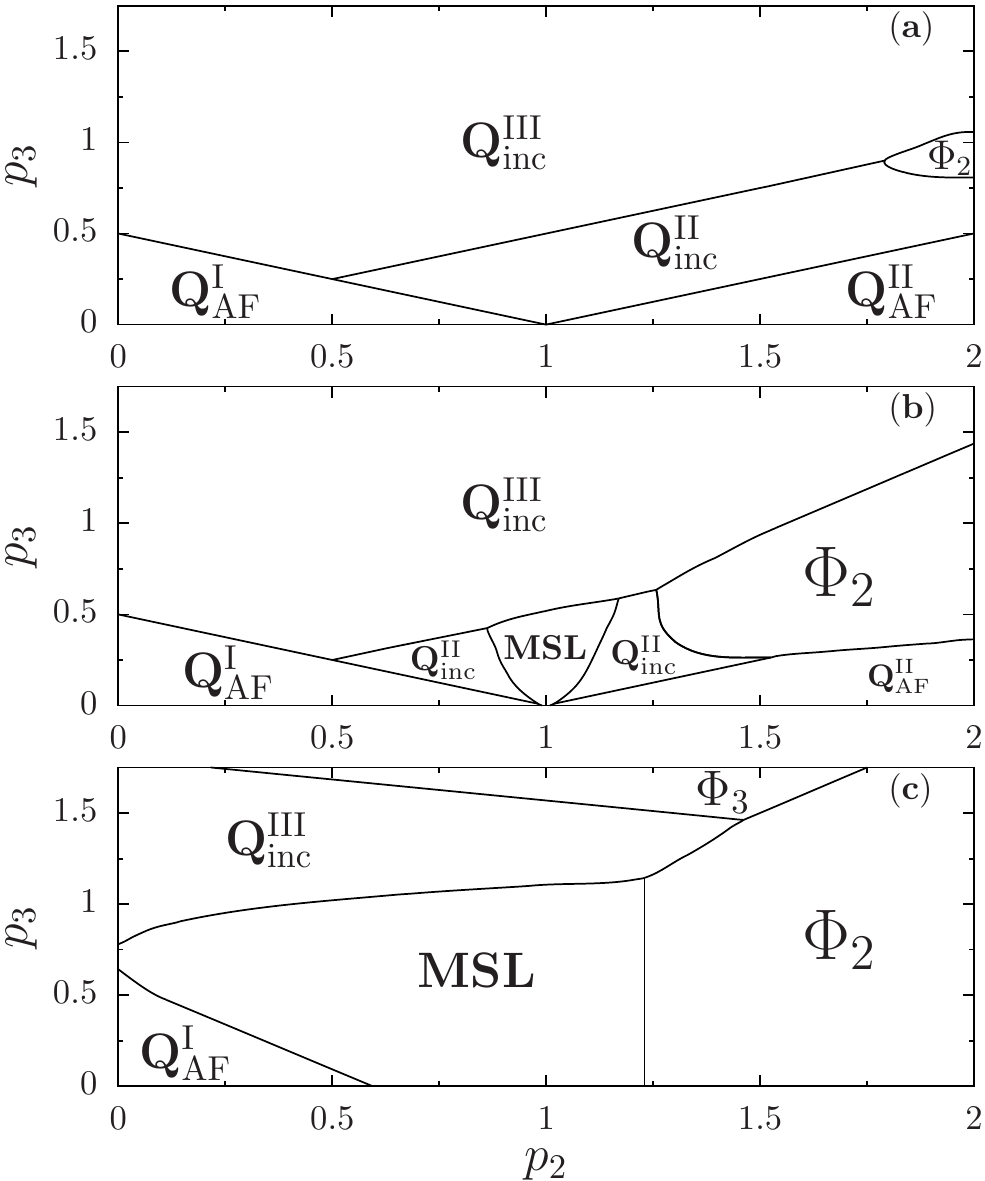}
\caption{Ground state phase diagram of the $J_1$-$J_2$-$J_3$ model in coordinates $(p_2,p_3)$ obtained for 
${\rm n}=4~{\rm (a)},~5~{\rm (b)},~7~{\rm (c)}$. 
\label{figGeneralPhasediag}}
\end{figure}

{\it Phase diagram.---}For ${\rm n}\le 3$ we find  purely AF ground states, and for 
${\rm n}\ge 10$ the most stable states are SL characterized by finite values of 
either $\Phi_{\rm M}$, $\Phi_2$, or $\Phi_3$. 
At mean-field level, these three SL parameters do not coexist. Furthermore, we remark that the transition 
between the MSL and the $\Phi_2$-dominated SL phases is first order. 
Beyond mean-field, we expect that only the SL critical temperature associated with a non-zero $\Phi_{\rm M}$ still corresponds to a phase transition signaled by the translation symmetry breaking.  
For $4\le {\rm n}\le 9$ we find a rich 
phase diagram exhibiting AF/SL quantum phase transitions that are controlled by $J_1$-$J_2$-$J_3$ parameters, 
as illustrated on figure~\ref{figGeneralPhasediag}. 
Increasing ${\rm n}$, AF-SL instability shows up first within the ${\bf Q}_{\rm inc}^{\rm II}$ phase. 
Furthermore, comparing the large-$S$ phase diagram 
(fig.~\ref{fig:classical}) with the one obtained for ${\rm n}=4$ (fig.~\ref{figGeneralPhasediag}), one observes 
that the $S_c=1/2$ spin-wave instability is located in the same region where the $\Phi_2$ dominant state becomes stabilized. Also, beyond the specificity of the associated order parameters, 
figure~\ref{figGeneralPhasediag} indicates that the SL instability "propagates" from large-$p_2$ ($\Phi_2$) to smaller-$p_2$ (MSL) areas if we increase the value of ${\rm n}$. 

{\it MSL phase for $S=1/2$.---}An interesting feature also appears for the MSL solution: with a relatively high numerical accuracy 
the modulation field $\Phi_{\rm M}$ is found to be always equal to the homogeneous field $\Phi_1$. 
This leads to a very extreme situation for the inter-layer field 
$\varphi_{\vri\vrri}^{1}=\frac{1}{2}[\Phi_1\pm\Phi_{\rm M}]$ which vanishes on half of the bonds while it keeps 
the finite value $\Phi_1=\Phi_{\rm M}$ on the other bonds. 
Introducing the probability $p_{\bf RR'}^{singlet}$ that a given bond ${\bf RR'}$ forms a singlet, 
the formation of the MSL state can be interpreted here as follows: 
first, for all the inter-layer bonds such that ${\bf Q}_{\rm AF}^{\rm I}\cdot ({\bf R}+{\bf R'})/2=\pi/2$, 
the interaction terms are effectively decoupled at mean-field level, leading to $p_{\bf RR'}^{singlet}=1/4$ and 
$\langle \vec{S}_{\bf R}\cdot\vec{S}_{\bf R'}\rangle=0$ . Then the SL with 
$\langle \vec{S}_{\bf R}\cdot\vec{S}_{\bf R'}\rangle\neq 0$ is formed on the other inter-layer bonds, with 
${\bf Q}_{\rm AF}^{\rm I}\cdot ({\bf R}+{\bf R'})/2=- \pi/2$, that remain effectively coupled. Using the numerical value 
$\Phi_1=\Phi_{\rm M}\approx 0.45$ computed at $T=0$ in the MSL, we find that $p_{\bf RR'}^{singlet}\approx 0.60$
on these effectively coupled bonds. This value has to be compared with the value $ln(2)\approx 0.69$ 
that is predicted for a one-dimensional 
Heisenberg chain using exact methods~\cite{Bethe1931, Schollwock2005}. We may thus interpret the MSL as a crystal of interacting 
filaments formed by the connected effectively coupled bonds. In this picture, spin excitations are deconfined fermions moving along these filaments. This may generalize the usual concept of valence bond crystal~\cite{Frustratedmagnetismbook} where localized spin $1$ excitations 
correspond to confined fermions. 

{\it Discussion.---}Here we considered a model with only localized spins. However we know from previous works on heavy-fermions and 
cuprates that charge fluctuations play a crucial role in destabilizing AF states. 
In the context of cuprates, the AF phase of the insulating parent compounds corresponds to $\vqaf^{\rm II}$. The SL phase introduced by Anderson 
{\it et al.}~\cite{Fazekas1974,Anderson1987, Baskaran1987, Rice1993, Wen1996} corresponds to the homogeneous 
correlated state associated here with $\Phi_2\neq 0$. 
One crucial specificity of this SL-scenario for superconductivity in cuprates relies on the two-dimensionality of the system. 
Stabilizing a SL state in 3D is commonly thought to be more tricky in view of the fact that the corresponding linear SW  
correction remains finite within a large $S$ approach. 
Nevertheless, we have shown how frustration in the BCT-lattice can enhance the critical value 
$S_c$ that, in some sense, characterizes the weakness of AF order against SW fluctuations. In connection with this weakening, we have identified various SL-phases that can be stabilized when ${\rm n}$ is larger than a relatively small critical value. 
This opens completely new perspectives for the realization of unconventional electronic quantum orderings in 3D. 
In particular, these results suggest that the BCT-lattice structure can play a central role in crystaline materials 
like the 122 and some cuprates in spite of numerous phases emerging in these systems.  
Indeed, the very rich phase diagram depicted in figures~\ref{fig:classical} and~\ref{figGeneralPhasediag} could provide a unifying framework for understanding and analyzing the intersite correlations in these compounds. The number 
${\rm n}$ may be considered as an effective parameter related to the electronic orbital degeneracy, that could be phenomenologically increased or decrerased by charge fluctuations or crystal field 
effects. Considering a given compound, ${\rm n}$ might also be effectively decreased by applying an external magnetic field. 
Similarly, a tuning of the  model parameters $p_2$ and $p_3$ may phenomenologically account for some effects of applying pressure~\cite{Thomas2013}. 
For example, this scenario could explain two different AF instabilities of the HO phase that are observed experimentally in 
URu$_2$Si$_2$: assuming that HO is a MSL order, and applying a finite pressure this corresponds to increasing $J_1$ for 
a fixed ${\rm n}$ (see fig.~\ref{figGeneralPhasediag}), leading to a commensurate AF instability characterized by ${\bf Q}_{\rm AF}^{\rm I}$ as observed experimentally~\cite{Palstra1985, Mydosh2011}. 
Alternatively, applying a magnetic field without pressure corresponds to lowering the effective value of ${\rm n}$ 
for fixed $p_2$ and $p_3$: the MSL (HO) is destabilized to an incommensurate AF. Interestingly, using different numerical values of $J_1$, $J_2$, and $J_3$ obtained from different fits of Inelastic Neutron Scattering datas, our scenario predicts 
an instability from MSL to ${\bf Q}_{\rm inc}^{\rm III}=({\Upsilon_3},0,0)$ with $\Upsilon_3\approx 0.69$ (from~\cite{Broholm1991}), $0.66$ (from~\cite{Kusunose2012}), $0.69$ (from~\cite{SUGIYAMA1990}) 
and $0.65$ (from~\cite{BourdarotHDR}). 
This scenario could be tested experimentally since it predicts that the AF order 
${\bf Q}_{\rm inc}^{\rm III}$ could be continuously tuned to ${\bf Q}_{\rm AF}^{\rm I}$ by applying pressure on 
URu$_2$Si$_2$ under a high magnetic field.

Invoking the SL instabilities described in this letter we may also generalize to 3D systems the spin-fluctuation pairing mechanism that was proposed in terms of gauge transformations in~\cite{Lee2006,Wenbook}. Here, the link between the BCT lattice structure and the superconducting order parameter is natural. It can be tested experimentally since we predict that the symmetries of the resulting superconducting order parameters will result from the point group symmetries of the SL, which may correspond to inplane pairing related to $\Phi_2$ or $\Phi_3$ or fully 3D pairing associated with $\Phi_1$. 
This SL mechanism driven by frustration on the BCT lattice may also be tested for the heavy-fermion 
superconductors CeRu$_2$Si$_2$ and CePd$_2$Si$_2$, but in these systems valence fluctuation effects need to be 
carefully included. 
The possible formation of a 
MSL could also give rise to a commensurately ordered pairing that would break the BCT-symmetry down to simple tetragonal. Such a modulated pairing unconventional scenario could be tested with the superconducting instability observed in URu$_2$Si$_2$ inside the HO phase. Alternatively, even if the chiral SL order was found here to be less stable than the MSL, an opposite result could occur by including charge fluctuations. 

\begin{acknowledgments}
We acknowledge the financial support of Capes-Cofecub Ph 743-12. CT is {\it bolsista Capes}.
This research was also supported in part by the Brazilian Ministry of Science, Technology and Innovation (MCTI) and the Conselho Nacional de Desenvolvimento Cient\'ifico e Tecnol\'ogico (CNPq). Research carried out with the aid of the Computer System of High Performance of the International Institute of Physics-UFRN, Natal, Brazil. The authors are gratefull to Fr\'ed\'eric Bourdarot for usefull discussions. 
\end{acknowledgments}

\bibliography{biblio_SpinLiquid_BCT}

\end{document}